\documentclass[aps,prd,superscriptaddress,onecolumn,showpacs,showkeys]{revtex4}
\usepackage{amssymb,amsmath,epsfig}

\usepackage{eurosym}
\usepackage{amsfonts}
\usepackage{array}
\usepackage{amsthm}
\usepackage{bm}
\usepackage{palatino}
\usepackage{mathpazo}
\usepackage{amssymb}
\usepackage{eurosym}
\usepackage{amsmath}
\usepackage{epsfig}
\usepackage{graphics}
\usepackage{changes}
\usepackage{color}
\usepackage{graphicx}

\begin{document}

\title{Effect of the dilaton field and plasma medium on deflection angle by black holes in Einstein-Maxwell-Dilaton-Axion theory}

\author{W. Javed} \email{wajiha.javed@ue.edu.pk;
wajihajaved84@yahoo.com}
\affiliation{Department of Mathematics, University of Education,\\
Township, Lahore-54590, Pakistan.}

\author{R. Babar}
\email{rimsha.babar10@gmail.com}
\affiliation{Department of Mathematics, University of Education,\\
Township, Lahore-54590, Pakistan.}

\author{A. \"{O}vg\"{u}n}
\email{ali.ovgun@pucv.cl}
\homepage[]{https://www.aovgun.com}
\affiliation{Instituto de F\'{\i}sica, Pontificia Universidad Cat\'olica de Valpara\'{\i}%
so, Casilla 4950, Valpara\'{\i}so, Chile}
\affiliation{Physics Department, Faculty of Arts and Sciences, Eastern Mediterranean
University, Famagusta, North Cyprus, via Mersin 10, Turkey}

\date{\today}

\begin{abstract}
In this paper, we argue that one can calculate the weak deflection angle of light in the background of Einstein-Maxwell-Dilaton-Axion black hole using the Gauss-Bonnet theorem. To support this, the optical geometry of the black hole with the Gibbons-Werner method are used to obtain the deflection angle of light in the weak field limits. Moreover, we investigate the effect of a plasma medium on deflection of light for a given black hole.
Because of dilaton and axion are one of the candidate of the dark matter, it can give us a hint on observation of dark matter which is supported by the black hole. Hence we demonstrate the observational viability via showing the effect of the dark matter on the weak deflection angle of light.
\end{abstract}


\maketitle

\section{Introduction}
Since the first photo of the Messier 87 black hole by Event Horizon Telescope, studying on black hole has been getting quite a lot of attention lately \cite{Akiyama:2019cqa}. On the other hand, the Laser Interferometer Gravitational-Wave Observatory (LIGO) has spotted the signs of yet another enormous collision in space, and this one seems to be between a black hole and a neutron star \cite{Authors:2019qbw}. As fascinating as it is mysterious, dark matter is one of the greatest enigmas of astrophysics and cosmology. Recently the existence of dark matter are shown by SISSA which disproves the empirical relations in support of alternative theories \cite{DiPaolo:2018mae}. Moreover, the ongoing estimations by the WMAP mission \cite{R1} have set up the relative plenitude of dark and baryonic matter in our Universe with incredible accuracy. Moreover, most of the content within the Universe is in the frame of cold dark matter, whose composition is still obscure. In order to solve the enigma of the dark matter the most promising theoretical particle \emph{axion} is proposed \cite{R2, R3}. This particle was first taken into consideration in $1977$ by way of R. Peccei and H. Quinn \cite{R4} in their notion to fathom the strong-CP problem in QCD theory. In spite of the fact that the first PQ axion is by presently prohibited, other axion models are still reasonable.  In literature \cite{R5,R6,R7,R8,R9} the hypothesis about axion that it might be the dark matter particle has been widely discussed. 

String theory is an auspicious aspirant for a consistent theory of quantum gravity and obviously the attributes of black holes (BHs) in string theory would be most intrigue. The string theory differs from general relativity (GR) due to the existence of scalar field which is known as dilaton field and causes the change in characteristics of BH geometries. In spite of dilaton, the axion has been accepted to be one of the strong candidate of the dark matter by numerous
physicists, and experiments to distinguish it have been effectively performed \cite{R10,R11,R12}. Although, the
identification of the dilaton has not all that effectively been performed up to now, despite its theoretical
significance.
It is important to mention that the dilaton produces the fifth force which can influence the Einstein's
gravity in a fundamental way \cite{R13,R14}.
Also, in cosmology it can assume the part of the inflation,
furthermore, can be an amazing competitor of the dark matter \cite{R10,R11,R12,R13,R14,R15}.
There exists solid tests and hypothetical prove for the presence of dark matter within the universe from gravitational lensing, galactic revolution curves and well-known inflationary models. The study will also offer new insights into understanding the nature of dilaton field and also axion dark matter.

It is a well known fact that the gravitational lensing is currently one of helpful instrument to search not just
for dark and massive objects, yet additionally wormholes and BHs.
Gravitational lensing is a specific effect of light deflection.
The gravitational bending of light by mass prompted the first exploratory confirmations of the general theory of relativity. 
The bending of light has theoretical significance, particularly for examining a null structure of a spacetime.
When a gravitating mass mutilates a spacetime, moreover, anything in it then gravitational lensing occurs. 
The trails observed via electromagnetic radiation from a star, universe, galaxy or other source are bent as properly.
In latest years, lensing has become an amazing test of numerous astrophysical and cosmological inquiries.
Strong lensing, or frameworks in which numerous pictures of a single source are recognizable or in which an Einstein ring is visible, can inform us regarding the Hubble constant and other cosmological parameters \cite{R16, R17}.
Factual estimations of lensing where the deflection of light is so weak to distinguish in a single background image, or weak lensing, gives an effective exploration of the matter distribution within the universe \cite{R18, R19}.
Weak lensing is an especially significant probe of dark matter \cite{R20,deLeon:2019qnp} and has been considered as an instrument to distinguish general relativity from different theories \cite{R21, R22}. 
In literature \cite{Werner:2012rc,Gibbons:2008rj,Gibbons:2015qja,Ishihara:2016vdc,Das:2016opi,Sakalli:2017ewb,Jusufi:2017lsl,Ono:2017pie,Jusufi:2017hed,Ishihara:2016sfv,Jusufi:2017vta,Goulart:2017iko,Jusufi:2017uhh,Arakida:2017hrm,Crisnejo:2018uyn,Jusufi:2018jof,Ovgun:2018fnk,Ovgun:2018ran,Ovgun:2018prw,Ono:2018ybw,Ovgun:2018oxk,Ovgun:2018tua,2,ao1,ao2,ao3,ao4}, a lot of researchers have studied the deflection angle of light for different types of BHs and wormholes by the following the formula of Gibbons and Werner which was proposed using the Gauss-Bonnet theorem:
\begin{equation}
\tilde{\delta}=-\int\int_{\mathcal{A}_{\infty}}\tilde{K}dS,
\end{equation}
where $\tilde{K}$ is the Gaussian curvature and $\mathcal{A}_{\infty}$ represents the infinite region of the surface. Recently, it has been studied the weak gravitational lensing by wormholes and calculated the deflection angle via naked singularities and compare their results \cite{R23}.

The main motivation of this paper is to study a conceivable extension of calculations of the bending angle of light for Einstein-Maxwell-Dilaton-Axion (EMDA) BH \cite{Sur:2005pm,Korunur:2012yc}. To do so, we calculate the deflection angle by null geodesic technique and then we will look at a relation among the
Gauss-Bonnet theorem and bending angle of light. After comparing these results, we investigate deflection angle in plasma medium for the given BH. The paper is organized as follows:
In Section \textbf{II}, we introduce the metric for EMDA BH. In In Section \textbf{III}, we investigate the deflection angle by geodesic method. Section \textbf{IV} is based on the calculation of deflection angle of light by EMDA BH without plasma medium in weak field approximations.
Section \textbf{V}, provides the deflection angle for EMDA BH in presence of plasma medium.
Finally, Section \textbf{VI} is devoted for the main results of this paper.

\section{
Metric Tensor of Black Hole in Einstein-Maxwell-Dilaton-Axion Theory}
The EMDA black hole metric with spherical coordinates is defined as \cite{Korunur:2012yc}
\begin{equation}
ds^2=-\tilde{A}(r)dt^2+\frac{dr^2}{\tilde{B}(r)}+\tilde{C}(r)\left[d\vartheta^{2}+\sin^{2}
\vartheta d\varphi^{2}\right],\label{a3}
\end{equation}

where
\begin{eqnarray}
\tilde{A}(r)&=&\tilde{B}(r)=\left(1-\frac{2 \tilde{M}}{r-2 r_{0}}\right) \text { and } ,~~~\tilde{C}(r)
=\left(r-2 r_{0}\right)^2,
\end{eqnarray}
here $\tilde{M}$ represents the mass of BH and $r_0$ is the parameter of dilaton-axion field where $r_0\approx a^2e^{-a\varphi}Q^2/4m_{\circ}$, $a$ is the constant free parameter and $Q$ represents the charge of BH and $m_{\circ}$ shows the relation $m_{\circ}\approx m+r_{\circ}$.

\section{Calculation of Deflection Angle by Geodesic Method}
In order to find the geodesics of EMDA BH,
the Lagrangian $\mathcal{\tilde{L}}$ can be given by the metric (\ref{a3}) as follows
\begin{eqnarray}
2\mathcal{\tilde{L}}&=&-\left(1-\frac{2\tilde{M}}{r(s)-2 r_{0}(s)}\right)\dot{t}^2(s)
+\left(1-\frac{2\tilde{M}}{r(s)-2 r_{0}(s)}\right)^{-1}\dot{r}^2(s)
+\left(r(s)-2 r_{0}(s)\right)^2\left(\dot{\vartheta}^2(s)
+\sin^2\vartheta(s)\dot{\varphi}^2(s)\right).~~~~~~\label{a26}
\end{eqnarray}

By considering $2\mathcal{\tilde{L}}=0$ for photons, we have the two constants
of motion of the geodesics at the equatorial plane
$\vartheta=\frac{\pi}{2}$ as follows
\begin{eqnarray}
\mathcal{\tilde{P}}_\varphi&=&\frac{\partial\mathcal{\tilde{L}}}{\partial\dot{\varphi}}
=2\left(r(s)-2 r_{0}\right)^2\dot{\varphi}(s)=\tilde{\ell},\label{a27}\\
\mathcal{\tilde{P}}_t&=&\frac{\partial\mathcal{\tilde{L}}}{\partial\dot{t}}
=-2\left(1-\frac{2\tilde{M}}{r(s)-2 r_{0}}\right)\dot{t}(s)=-\tilde{\epsilon}.\label{a28}
\end{eqnarray}

Furthermore, we consider a new variable $\xi(\varphi)$, which is associated
to the old radial coordinate as $r=\frac{1}{\xi(\varphi)}$, which leads to the identity
\begin{equation}
\frac{\dot{r}}{\dot{\varphi}}=\frac{dr}{d\varphi}=-\frac{1}{\xi^2}\frac{d\xi}{d\varphi}.\label{a30}
\end{equation}
For the sake of simplicity, we use the metric conditions $\tilde{\epsilon}=1$ \& $\tilde{\ell}=b$ (note that $b$ is impact parameter)
for Eqs. (\ref{a26})-(\ref{a30}) and after some
algebraic manipulations we have the following relation
\begin{equation}
\frac{1}{\xi^4}\left(\frac{d\xi}{d\varphi}\right)^2\left(1-\frac{2\tilde{M}}{1/\xi-2 r_{0}}\right)^{-1}-\frac{
(1/\xi-2r_{0})^4}{b^2}\left(1-\frac{2\tilde{M}}{1/\xi-2 r_{0}}\right)^{-1}+(1/\xi-2r_{0})^2=0.
\end{equation}
The above equation implies
\begin{equation}
\left(\frac{d\varphi}{d\xi}\right)=\frac{\pm b}{\sqrt{1-2r_{0}\xi}}\frac{1}{\sqrt{1+(-6\xi+2b^2\xi^3)r_0+12r_{0}^2\xi^2-8r_{0}^3\xi^3-b^2(\xi^2-2\tilde{M}\xi^3)}},\label{a28}
\end{equation}

In order to derive the solution of differential Eq. (\ref{a28}),
we use the following relation \cite{A3}
\begin{equation}
\Delta\varphi=\pi+\tilde{\delta},
\end{equation}
where $\tilde{\delta}$ represents the deflection angle. The deflection
angle can be obtained by following the same procedure of Ref. \cite{A4}
\begin{equation}
\tilde{\delta}=2|\varphi_{\xi=1/b}-\varphi_{\xi=0}|- \pi,
\end{equation}
and
\begin{eqnarray}
\tilde{\delta}&=&\mid\int_{0}^{1/b}\left(\frac{d\varphi}{d\xi}\right)d\xi\mid-\pi\nonumber\\
&=&\mid\int_{0}^{1/b}\frac{ b}{\sqrt{1-2r_{0}\xi}}\frac{1}{\sqrt{1+(-6\xi+2b^2\xi^3)r_0+12r_{0}^2\xi^2-8r_{0}^3\xi^3-b^2(\xi^2-2\tilde{M}\xi^3)}}\mid-\pi.
\end{eqnarray}

Here we neglect all the terms of $r_{0}^2$ and $r_{0}^3$,

\begin{equation}
\tilde{\delta}=\mid\int_{0}^{1/b}\frac{b(1+r_{0}\xi)}{\sqrt{1+(-6\xi+2b^2\xi^3)r_0-b^2(\xi^2-2\tilde{M}\xi^3)}}d\xi\mid-\pi.
\end{equation}
Then we proceed to introduce new variable $y = 1/b$
and expand in Taylor series around $y$. After we evaluate
the integral for leading order terms of $\tilde{M}$ and $r_0$, the deflection angle in the weak deflection
limit approximation is found to be

\begin{equation}
\tilde{\delta}\simeq{\frac {4\tilde{M}}{b}}+\,{\frac {3{\it r_0}\,\tilde{M}\pi}{2{b}^{2}}}.\label{b1}
\end{equation}
The deflection angle depends upon mass $\tilde{M}$, parameter of $r_0$ and impact parameter $b$. Increasing the value of dilaton-axion $r_0$ parameter, increases the deflection angle in weak field limits.
\section{Calculation of Deflection Angle by Gauss-Bonnet Theorem}

In order to find the null geodesics $(ds^2=0)$, the  metric can be written in simplified form as follows \cite{2}
\begin{equation}
dt^2=\frac{dr^2}{\tilde{A}(r)^2}+\frac{\tilde{C}(r)d\varphi^2}{\tilde{A}(r)}.
\end{equation}.
We consider a new coordinate $r^\ast$ which satisfies the metric tensor $g_{\mu\nu}$ as
\begin{equation}
dt^2=g_{\mu\nu}dx^{\mu}dx^{\nu}=dr^{\ast}{^2}+f^2(r^{\ast})d\varphi^2,
\end{equation}
where
\begin{equation}
f(r^\ast)=\sqrt{\frac{\tilde{C}(r)}{\tilde{A}(r)}}
\end{equation}
In order to consider the Gauss-Bonnet theorem, firstly, we calculate
the Gaussian curvature $\tilde{K}$ of the optical spacetime which
is an intrinsic property of spacetime. The Gaussian curvature can be calculated as \cite{Ovgun:2018prw}
\begin{equation}
\tilde{K}=-\frac{R_{r\varphi r\varphi}}{\det g_{r\varphi}}=-\frac{1}{f(r^{\ast})}\frac{d^2f(r^{\ast})}{dr^{\ast}{^2}}=-2\,{\frac {\tilde{M}}{{r}^{3}}}+3\,{\frac {{\tilde{M}}^{2}}{{r}^{4}}}+ \left( -6\,{
\frac {\tilde{M}}{{r}^{4}}}+12\,{\frac {{\tilde{M}}^{2}}{{r}^{5}}} \right) {\it r_0}.
\label{z1}
\end{equation}
The Gaussian curvature depends upon the mass of the BH.

The Gauss-Bonnet theorem (GBT) with boundary $\partial\mathcal
{\tilde{H}}_{R}=\tilde{\gamma}_{h}\cup\mathcal{C}_{R}$, can be defined as \cite{Gibbons:2008rj}
\begin{equation}
\iint_{\mathcal{\tilde{H}}_{R}}\tilde{K}d\tilde{\sigma}+\oint_{\partial\mathcal
{\tilde{H}}_R}\tilde{\kappa}dt+\sum_i\tilde{\theta}_{i}=2\pi\tilde{\chi}(\mathcal
{\tilde{H}}_{R}),
\end{equation}
where $\tilde{\kappa}$ denotes the geodesic curvature, while $\tilde{\theta}_{i}$
represents the exterior angle at the $ith$ vertex and we consider non-singular domain $\mathcal{\tilde{H}}_{R}$ outside of the light ray with Euler characteristic $\tilde{\chi}(\mathcal{\tilde{H}}_{R})=1$ \cite{Ovgun:2018prw}. When, we consider
$\mathcal{R}\rightarrow\infty$, then the two jump angles $(\tilde{\theta}_
\mathcal{O}, \tilde{\theta}_\mathcal{S})$ yield $\frac{\pi}{2}$ and also, if we
consider the total sum of jump angles at observer $\mathcal{O}$ and
source $\mathcal{S}$, we get $(\tilde{\theta}_
\mathcal{O}+\tilde{\theta}_\mathcal{S})\rightarrow\pi$ \cite{Gibbons:2008rj}.
So, the GBT can be rewritten as follows
\begin{equation}
\iint_{\mathcal{\tilde{H}}_{R}}\tilde{K}d\tilde{\sigma}+\oint_{\mathcal
{C}_R}\tilde{\kappa}dt =^{\mathcal{R}\rightarrow\infty}\iint_
{\mathcal{\tilde{H}_{\infty}}}\mathcal{\tilde{K}}d\tilde{\sigma}+\int^{\pi+\tilde{\alpha}}_0d\phi=\pi. \label{gbtt}
\end{equation}
Now, in order to compute the geodesic curvature $\tilde{\kappa}$,
we follow $\tilde{\kappa}(\tilde{\gamma}_{h})=0$, since
$\tilde{\gamma}_{h}$ represents the geodesic. Therefore, we get
\begin{equation}
\tilde{\kappa}(C_R)=|\nabla_{\dot{C}_
R}\dot{C}_R|,\label{a5}
\end{equation}
in which we can choose $C_R:=r(\phi)=R=\texttt{const}$, where $C_R$ represents the circle segment of coordinate radius $R$.
Thus the non-zero radial part can be calculated by
\begin{equation}
\left(\nabla_{\dot{C}_R}\dot{C}_{R}\right)^r
=\dot{C}_R^\phi(\partial_\phi\dot{C}_R^r)+
\tilde{\Gamma}^{r(opt)}_{\phi\phi}(\dot{C}_R^\phi)^2,
\end{equation}
where $\tilde{\Gamma}^{r(opt)}_{\phi\phi}$ is the Christoffel symbol related
to the optical geometry. It is clear that the first term in the above equation
vanishes the second term $(\dot{C}_R^\phi)^2=1/f^2(r^\ast)$ and $\tilde{\Gamma}^{r(opt)}_{\phi\phi}=f(r^\ast)f^{'}(r^\ast)$. When $R\rightarrow\infty$, then the geodesic
curvature $\tilde{\kappa}$ and $dt$ becomes
\begin{eqnarray}
\lim_{R\rightarrow\infty}\tilde{\kappa}
(C_R)&=&\lim_{R\rightarrow\infty}
|\nabla_{\dot{C}_R}\dot{C}_R|,\nonumber\\
&\rightarrow&\frac{1}{R}.\label{a6}
\end{eqnarray}
Also
\begin{eqnarray}
\lim_{R\rightarrow\infty}dt&=&\lim_{R\rightarrow\infty}
\left(\frac{\tilde{C}}{\tilde{A}}\right)^{1/2}d\phi,\nonumber\\
&\rightarrow& (R-2r_0)d\phi.\label{a7}
\end{eqnarray}
By combining Eqs. (\ref{a6}) \& (\ref{a7}), we can get
$\tilde{\kappa}(C_R)dt \approx Rd\phi$.

Now we discuss the deflection angle in weak field limits. It is a well
known fact that in weak field regions the light rays persues a straight
line approximation, therefore we consider the condition of $r=\frac{b}{\sin\varphi}$ zero order.
By using Eq. (\ref{z1}), Eq. (\ref{gbtt}) and Eq. (\ref{a7}), the formula for deflection angle is obtain as follows
\begin{equation}
\tilde{\delta}=
-\int_{0}^{\pi}\int_{\frac{b}{\sin\varphi}}^{\infty}\tilde{K}d\tilde{\sigma}.
\end{equation}
After using the above relation the deflection angle $\tilde{\delta}$ only for leading order of $\tilde{M}$ can be calculated as
\begin{equation}
\tilde{\delta}\simeq{\frac {4\tilde{M}}{b}}+\,{\frac {3{\it r_0}\,\tilde{M}\pi}{2{b}^{
2}}}.\label{d1}
\end{equation}
The deflection angle depends upon mass $\tilde{M}$, parameter of $r_0$ and impact parameter $b$. Increasing the value of dilaton-axion $r_0$ parameter, increases the deflection angle in weak field limits.
\section{Weak gravitational lensing of EMDA BH in plasma medium}
In this section, we study the effect of plasma medium on the weak gravitational lensing by EMDA BH.
The refractive index for the EMDA BH is given as \cite{Crisnejo:2018uyn}
\begin{equation}
\tilde{n}(r)=\sqrt{1-\frac{\tilde{\omega}_e^2}{\tilde{\omega}_{\infty}^2}\left(1-\frac{2 \tilde{M}}{r-2 r_{0}}\right)}.\label{zr}
\end{equation}
where $\tilde{\omega}_e$ and $\tilde{\omega}_{\infty}$ denotes the electron and photon plasma frequencies, respectively.

In order to study the application of Gauss-Bonnet theorem in the determination of bending angle which followed by Gibbons and Werner \cite{Gibbons:2008rj}, we consider two dimensional Riemannian manifold $(\mathcal{M}^{opt}, g_{mn}^{opt})$ with the optical metric $g_{mn}^{opt}=-\frac{n^2}{g_{00}}g_{mn}$, so the corresponding optical metric is defined as follows
\begin{equation}
d\tilde{\sigma}^2=g_{mn}^{opt}dx^mdx^n=\frac{\tilde{n}^2(r)}
{\tilde{A}(r)}\left(\frac{1}{\tilde{A}(r)}dr^2+\tilde{C}(r)d\varphi^2\right),~~~\textmd{where}~~~m, n=1,2,3...,\label{z3}
\end{equation}
This metric preserves the angle between two curves at a given point and
is conformally related to the metric (\ref{a3}), when we choose spatial section $t=constant,~\theta=\pi/2$.
By using Eq. (\ref{zr}) into Eq. (\ref{z3}), the optical metric gets the form
\begin{equation}
d\tilde{\sigma}^2=\frac{r(\tilde{\omega}_{\infty}^2-\tilde{\omega}_e^2)-2r_0(\tilde{\omega}_{\infty}^2-\tilde{\omega}_e^2)+2\tilde{M}\tilde{\omega}_e^2}{(r-2r_0-2\tilde{M})\tilde{\omega}_{\infty}^2}\left(\frac{dr^2}{1-\frac{2\tilde{M}}{r-2r_0}}+r^2d\varphi^2\right).
\end{equation}
The Gaussian curvature can be calculated as follows
\begin{equation}
\tilde{K}=-\frac{R_{r\varphi r\varphi}(g^{opt})}{\det (g^{opt})}={\frac {\tilde{M} \left( {\omega_e}^{2}r-2\,{\omega_{\infty}}^{2}r+4
\,{\it r_0}\,{\omega_e}^{2}-6\,{\it r_0}\,{\omega _{\infty}}^{2}
 \right) {\omega_{\infty}}^{2}}{ \left( {\omega_e}^{2}-{
\omega_{\infty}}^{2} \right) ^{2}{r}^{4}}}.
\end{equation}
Moreover, we have
\begin{equation}
\frac{d\tilde{\sigma}}{d\varphi}\mid_{\mathcal{C}_R}=n(R)
\left(\frac{\tilde{C}(R)}{\tilde{A}(R)}\right)^{1/2},
\end{equation}
which implies
\begin{equation}
\lim_{R\rightarrow\infty}\tilde{\kappa}_{g}\frac{d\tilde{\sigma}}{d\varphi}
\mid_{\mathcal{C}_R}\approx \frac{1}{R}.
\end{equation}
By taking the straight line approximation $r=\frac{b}{\sin\varphi}$ and
for the limit $R\rightarrow\infty$, the Gauss Bonnet theorem gets the form
\begin{equation}
\lim_{R\rightarrow\infty}\int_{0}^{\pi+\tilde{\delta}}\left[\tilde{\kappa}_{g}\frac{d\tilde{\sigma}}{d\varphi}
\right]\mid_{C_{\mathcal{R}}}d\varphi=\pi-\int_{0}^{\pi}\int_{\frac{b}{\sin\varphi}}^{R}\tilde{K}d\tilde{\sigma}.
\end{equation}
So, the deflection angle is obtained in the form
\begin{equation}
\tilde{\delta}=3/2\,{\frac {\tilde{M}{\it r_0}\,\pi}{{b}^{2}}}+4\,{\frac {\tilde{M}}{b}}+5/4\,{\frac {
{\omega_e}^{2}\tilde{M}{\it r_0}\,\pi}{{\omega_{\infty}}^{2}{b}^{2}}}+
4\,{\frac {{\omega_e}^{2}\tilde{M}}{b{\omega_{\infty}}^{2}}}
+\mathcal{O}(\tilde{M}^2).\label{z5}
\end{equation}
The above equation indicates the photon rays motion in
a medium of homogeneous plasma. It is worth to note that, if we
neglect the plasma effects, i.e., $\frac{\omega_{e}^2}{\omega_{\infty}^2}\rightarrow0$,
then the Eq. (\ref{z5}) reduces into Eq. (\ref{d1}), so we can observe that the plasma effects can be removed. Moreover, the dilaton-axion paramter $r_0$ increases the deflection angle in a medium of homogeneous plasma in weak field limits.

\section{Conclusion}
In this work, we have studied the deflection angle for the EMDA BH.
To do so, by considering the null geodesic method as well as
new geometric techniques (Gauss-Bonnet theorem and optical geometry) established by Gibbons and Werner, we
have calculated the deflection angle for EMDA BH.
It is also important to note that the deflection of a
light ray is calculated outside of the lensing area
which shows that the gravitational lensing effect
is a global and even topological effect, i.e., there are more than one light ray converging between the source and observer.
Hence the deflection angle of photon is calculated as follows:
\begin{equation}
\tilde{\delta}\simeq{\frac {4\tilde{M}}{b}}+\,{\frac {3{\it r_0}\,\tilde{M}\pi}{2{b}^{
2}}}.\label{d111}
\end{equation}

On the other hand, the deflection angle of  photon is obtained in the form:
\begin{equation}
\tilde{\delta}=3/2\,{\frac {\tilde{M}{\it r_0}\,\pi}{{b}^{2}}}+4\,{\frac {\tilde{M}}{b}}+5/4\,{\frac {
{\omega_e}^{2}\tilde{M}{\it r_0}\,\pi}{{\omega_{\infty}}^{2}{b}^{2}}}+
4\,{\frac {{\omega_e}^{2}\tilde{M}}{b{\omega_{\infty}}^{2}}}
+\mathcal{O}(\tilde{M}^2).\label{z544}
\end{equation}
Hence, we show that the dilaton-axion paramter $r_0$ increases the deflection angle in a medium of homogeneous plasma in weak field limits.
\acknowledgments
This work was supported by Comisi{\'o}n Nacional de Ciencias y Tecnolog{\'i}a of Chile through FONDECYT Grant $N^\mathrm{o}$ 3170035 (A. {\"O}.).


\begin{thebibliography}{99}

\bibitem{Akiyama:2019cqa} 
  K.~Akiyama {\it et al.} [Event Horizon Telescope Collaboration],
  Astrophys.\ J.\  {\bf 875}, no. 1, L1 (2019).
  
  \bibitem{Authors:2019qbw} 
  A.~Authors [LIGO Scientific and Virgo Collaborations],
   Phys.\ Rev.\ Lett.\  {\bf 123}, 161102 (2019).
  
\bibitem{DiPaolo:2018mae} 
  C.~Di Paolo, P.~Salucci and J.~P.~Fontaine,
  Astrophys.\ J.\  {\bf 873}, no. 2, 106 (2019)
  
\bibitem{R1} E. Komatsu et al., Astrophys. J. Suppl. \textbf{180}, 330 (2009).

\bibitem{R2} S. Weinberg, Phys. Rev. Lett. \textbf{40}, 223 (1978).

\bibitem{R3} F. Wilczek, Phys. Rev. Lett. \textbf{40}, 279 (1978).

\bibitem{R4} R. D. Peccei, H. R. Quinn, Phys. Rev. Lett. \textbf{38}, 1440 (1977).

\bibitem{R5} J. E. Kim, Phys. Rev. Lett. \textbf{43}, 103 (1979).

\bibitem{R6} M. A. Shifman, A. I. Vainshtein and V. I. Zakharov, Nucl. Phys. \textbf{B 166}, 493 (1980).

\bibitem{R7} M. Dine, W. Fischler and M. Srednicki, Phys. Lett. \textbf{B 104}, 199 (1981).

\bibitem{R8} A. P. Zhitnitskii, Sov. J. Nucl. Phys. \textbf{31}, 260 (1980).

\bibitem{R9} J. Preskill, M. Wise and F. Wilczek, Phys. Lett. \textbf{B 120}, 127 (1983).

\bibitem{R10} P. Sikivie, Phys. Rev. Lett. 51, 16 (1983); ibid. Phys. Rev. \textbf{D 32}, 11 (1985).

\bibitem{R11}  P. Sikivie, D. Tanner, and Y. Wang, Phys. Rev. \textbf{D 50}, 4744 (1994).

\bibitem{R12} Y. M. Cho, J. Math. Phys. \textbf{16}, 2029 (1975); Y. M. Cho and P. G. O. Freund, Phys. Rev. \textbf{D 12}, 1711
(1975); Y. M. Cho and P. S. Jang, Phys. Rev.\textbf{ D 12}, 3138 (1975).

\bibitem{R13} Y. M. Cho, Phys. Rev. \textbf{D 35}, 2628 (1987); Phys. Lett. \textbf{B 199}, 358 (1987).

\bibitem{R14} Y. M. Cho and D. H. Park, Gen. Rel. Grav. \textbf{23}, 741 (1991).

\bibitem{R15}  Y. M. Cho and Y. Y. Keum, Class. Quant. Grav. \textbf{15}, 907 (1998).

\bibitem{R16} C. S. Kochanek, C. R. Keeton, and B. A. McLeod, Astrophys. J. \textbf{547}, 50 (2001).

\bibitem{R17} C. R. Keeton and C. S. Kochanek, Astrophys. J. \textbf{487}, 42 (1997).

\bibitem{R18} Y. Mellier, Ann. Rev. Astron. Astrophys. \textbf{37}, 127 (1999).

\bibitem{R19} M. Bartelmann and P. Schneider, Phys. Rep. \textbf{340}, 291 (2001).

\bibitem{R20} N. Kaiser and G. Squires, Astrophys. J. \textbf{404}, 441 (1993).

    \bibitem{deLeon:2019qnp} 
  K.~de Leon and I.~Vega,
  Phys.\ Rev.\ D {\bf 99}, no. 12, 124007 (2019).

\bibitem{R21} F. Schmidt, Phys. Rev. \textbf{D 78}, 043002 (2008).

\bibitem{R22} J. Guzik, B. Jain, and M. Takada, Phys. Rev. \textbf{D 81}, 023503 (2010).

  \bibitem{Werner:2012rc} 
  M.~C.~Werner,
  Gen.\ Rel.\ Grav.\  {\bf 44}, 3047 (2012).
  
   \bibitem{Gibbons:2008rj} 
  G.~W.~Gibbons and M.~C.~Werner,
  Class.\ Quant.\ Grav.\  {\bf 25}, 235009 (2008)

  \bibitem{Gibbons:2015qja} 
  G.~W.~Gibbons,
  Class.\ Quant.\ Grav.\  {\bf 33}, no. 2, 025004 (2016).
  
  \bibitem{Ishihara:2016vdc} 
  A.~Ishihara, Y.~Suzuki, T.~Ono, T.~Kitamura and H.~Asada,
  Phys.\ Rev.\ D {\bf 94}, no. 8, 084015 (2016).
  
  \bibitem{Das:2016opi} 
  P.~Das, R.~Sk and S.~Ghosh,
  Eur.\ Phys.\ J.\ C {\bf 77}, no. 11, 735 (2017).
  

  
  \bibitem{Sakalli:2017ewb} 
  I.~Sakalli and A.~\"{O}vg\"{u}n,
  EPL {\bf 118}, no. 6, 60006 (2017).
  
  
  
  \bibitem{Jusufi:2017lsl} 
  K.~Jusufi, M.~C.~Werner, A.~Banerjee and A.~\"{O}vg\"{u}n,
  Phys.\ Rev.\ D {\bf 95}, no. 10, 104012 (2017).
  
   
  
  
  \bibitem{Ono:2017pie} 
  T.~Ono, A.~Ishihara and H.~Asada,
  Phys.\ Rev.\ D {\bf 96}, no. 10, 104037 (2017).
  
  \bibitem{Jusufi:2017hed} 
  K.~Jusufi, I.~Sakalli and A.~\"{O}vg\"{u}n,
  Phys.\ Rev.\ D {\bf 96}, no. 2, 024040 (2017).
  
    \bibitem{Ishihara:2016sfv} 
  A.~Ishihara, Y.~Suzuki, T.~Ono and H.~Asada,
  Phys.\ Rev.\ D {\bf 95}, no. 4, 044017 (2017).
  

  
  
  \bibitem{Jusufi:2017vta} 
  K.~Jusufi, A.~\"{O}vg\"{u}n and A.~Banerjee,
  Phys.\ Rev.\ D {\bf 96}, no. 8, 084036 (2017).
  
  \bibitem{Goulart:2017iko} 
  P.~Goulart,
  Class.\ Quant.\ Grav.\  {\bf 35}, no. 2, 025012 (2018).
  
  

  
  \bibitem{Jusufi:2017uhh} 
  K.~Jusufi and A.~\"{O}vg\"{u}n,
  Phys.\ Rev.\ D {\bf 97}, no. 6, 064030 (2018).
  
  \bibitem{Arakida:2017hrm} 
  H.~Arakida,
  Gen.\ Rel.\ Grav.\  {\bf 50}, no. 5, 48 (2018).
  
 
  
   \bibitem{Crisnejo:2018uyn} 
  G.~Crisnejo and E.~Gallo,
  Phys.\ Rev.\ D {\bf 97}, no. 12, 124016 (2018).
  
  \bibitem{Jusufi:2018jof} 
  K.~Jusufi, A.~\"{O}vg\"{u}n, J.~Saavedra, Y.~Vasquez and P.~A.~Gonzalez,
  Phys.\ Rev.\ D {\bf 97}, no. 12, 124024 (2018).
  
  
       
  
  \bibitem{Ovgun:2018fnk} 
  A.~\"{O}vg\"{u}n,
  Phys.\ Rev.\ D {\bf 98}, no. 4, 044033 (2018).
  
  \bibitem{Ovgun:2018ran} 
  A.~\"{O}vg\"{u}n, K.~Jusufi and I.~Sakalli,
  Annals Phys.\  {\bf 399}, 193 (2018).
  
  

  \bibitem{Ovgun:2018prw} 
  A.~\"{O}vg\"{u}n, G.~Gyulchev and K.~Jusufi,
  Annals Phys.\  {\bf 406}, 152 (2019).
  
  \bibitem{Ono:2018ybw} 
  T.~Ono, A.~Ishihara and H.~Asada,
  Phys.\ Rev.\ D {\bf 98}, no. 4, 044047 (2018).
  
  \bibitem{Ovgun:2018oxk} 
  A.~\"{O}vg\"{u}n,
 Universe {\bf 5}, no. 5, 115 (2019).
  
  \bibitem{Ovgun:2018tua} 
  A.~\"{O}vg\"{u}n, I.~Sakalli and J.~Saavedra,
  JCAP {\bf 1810}, no. 10, 041 (2018).
  
 
\bibitem{2} A. \"{O}vg\"{u}n,  Phys.\ Rev.\ D {\bf 99}, no. 10, 104075 (2019).

\bibitem{ao1}W. Javed, J. Abbas, A. \"{O}vg\"{u}n,  Phys.\ Rev.\ D {\bf 100}, no. 4, 044052 (2019). 

 \bibitem{ao2}W. Javed, J. Abbas, A. \"{O}vg\"{u}n, Preprints 2019, 2019060124 (doi: 10.20944/preprints201906.0124.v1). 
 
 \bibitem{ao3}W. Javed, J. Abbas, A. \"{O}vg\"{u}n,  Eur.\ Phys.\ J.\ C {\bf 79}, no. 8, 694 (2019).

\bibitem{ao4} 
  Y.~Kumaran and A.~\"{O}vg\"{u}n,
  arXiv:1905.11710 [gr-qc].

  
\bibitem{R23}
 W. Javed, R. Babar and A.~\"{O}vg\"{u}n, Phys. Rev. \textbf{D 99}, 084012 (2019).

\bibitem{Sur:2005pm} 
  S.~Sur, S.~Das and S.~SenGupta,
  JHEP {\bf 0510}, 064 (2005)

  \bibitem{Korunur:2012yc} 
  M.~Korunur and I.~Acikgoz,
  Adv.\ High Energy Phys.\  {\bf 2012}, 301081 (2012).

\bibitem{1} A. Bhadra, Phys. Rev. \textbf{D67}, 103009(2003).

 \bibitem{A3} R. H. Boyer and R. W. Lindquist, J. Math. Phys. \textbf{8}, 265(1967).
 
 \bibitem{A4} S. Weinberg, \textit{Gravitation and Cosmology} (John Wiley \& Sons, New York, 1972).


\bibitem{Liang:2017vdd} 
  J.~Liang, Gen.\ Rel.\ Grav.\  {\bf 49}, no. 11, 137 (2017).
  

  

\end{thebibliography}
\end{document}